\documentclass[reprint,nofootinbib,amsmath,amssymb,aps]{revtex4-2}
\usepackage{aas_macros}
\usepackage{graphicx}
\usepackage{dcolumn}
\usepackage{bm}
\usepackage{amsmath}
\usepackage{amssymb}
\usepackage{booktabs}
\usepackage{txfonts}
\usepackage{rotating} 
\usepackage[T1]{fontenc}
\usepackage{amsbsy}
\usepackage{color}
\usepackage{xcolor}
\usepackage{subcaption}
\usepackage[colorlinks=true, allcolors=blue]{hyperref}
\usepackage[normalem]{ulem}
\usepackage{ragged2e} 
\usepackage{comment}

\newcommand{\be}{\begin{equation}}
\newcommand{\ee}{\end{equation}}

\begin{document}

\title{On the nature of the missing mass of galaxy clusters \\ in MOND: the view from gravitational lensing}
\author{Benoit Famaey}
\affiliation{Universit\'e de Strasbourg, CNRS, Observatoire astronomique de Strasbourg,\\ UMR 7550, F-67000 Strasbourg, France}
\author{Lorenzo Pizzuti}
\affiliation{Dipartimento di Fisica G. Occhialini, Universit\`a degli Studi di Milano Bicocca, \\ Piazza della Scienza 3, 20126 Milano, Italy}
\author{Ippocratis D. Saltas }
\affiliation{CEICO, Institute of Physics, Czech Academy of Sciences\\Na Slovance 2, 182 21 Praha 8, Czech Republic}

\date{\today}

\begin{abstract}
 
 Modified Newtonian Dynamics (MOND) has long been known to fail in galaxy clusters, implying a residual missing mass problem for clusters in this context. Here, using mass profiles derived from strong- and weak-lensing shear, as well as magnification data, for a sample of clusters from the CLASH survey, we characterize the density profile of this residual MOND missing mass in the central Mpc of galaxy clusters. In line with results obtained in the literature from the hydrostatic equilibrium of hot gas, we find that an inner constant density core and an outer power-law slope sharper than $-3.5$ provides a good description within $\sim 1$~Mpc. We also show that the data in the central parts of clusters can be even better represented by a `dark mass-follows-gas' profile with an exponential cut-off. Clusters in the sample display a remarkable uniformity for the missing-to-hot-gas density ratio in the inner parts, of order $\sim$10, as well as for the exponential cut-off radius, of order $\sim 400$~kpc. These lensing results can in principle serve as a crucible for relativistic theories of MOND in galaxy clusters, or for any other tentative hypothesis regarding the nature of the clusters residual missing mass in the MOND context.

\end{abstract}

\maketitle

\section{Introduction}
Dark matter (DM) plays a central role in the standard $\Lambda$CDM cosmological model. It is considered as a meaningful explanation for a large variety of discrepancies between astronomical measurements and the predictions of General Relativity (GR) based on baryonic matter alone. However, the inference of its exact distribution and properties depends crucially on the validity of GR on all scales considered. Various modifications of gravity (see, e.g., \cite{Lombriser2015}) can change the distribution of DM inferred from observational data. In principle, it is even possible to do fully away with DM in galaxies via an appropriate modification of Newtonian dynamics \citep[MOND,][]{Milgrom1983,Famaey2012,Famaey2025}. The idea is that Newtonian dynamics, the weak-field limit of GR, would break down at ultra-low accelerations typical of the outskirts of galaxies (below $a_0 \simeq 10^{-10}$~m/s$^2$) and that the acceleration would fall as $r^{-1}$ instead of $r^{-2}$ in this regime. Such a modification of gravity would imply, observationally, a strong correlation between the Newtonian gravity generated by the baryons and the total gravitational field in stellar systems. 

Actually, in rotationally-supported galaxies, observations indicate that the baryonic surface density of galaxies (and therefore the Newtonian gravity generated by the baryons) plays a central role in setting the inner shape of their observed rotation curves (supposedly set by the profile of their DM halo in the GR context), as if baryons were fully dominating even when they are not: the associated diversity of rotation curve shapes remains challenging to understand in the standard context \cite{Sancisi, Swaters, Oman, Ghari}, while it is a natural expectation in the context of MOND, see e.g. \citep{Famaey2012}. At the same time, the baryonic surface density plays absolutely no role in setting the tight relation \cite{McGaughBT,LelliBT, DesmondBT} between the baryonic mass $M_b$ and the asymptotic circular speed $V_f$, known as the baryonic Tully-Fisher relation (BTFR), $M_b \propto V_f^4$. The BTFR seems to be surprisingly valid at very large distances from galaxies, as inferred from weak lensing measurements from the Kilo-Degree Survey (KIDS) around both late-type and early-type galaxies \cite{Mistele}. This striking contrast between the central role of surface density in setting observationally the inner shape of galaxy rotation curves and its absence of role in setting the BTFR is not well understood in the standard context, but it is a prediction of MOND. More generally, the close relation observed between the gravitational acceleration generated by baryons and the total one, known as the Radial Acceleration Relation \cite{LelliRAR,Stiskalek}, is a direct prediction of MOND too. All this explains why MOND is still taken seriously today in the context of galaxy dynamics, although it is difficult to reconcile with the success of our current standard model of cosmology. MOND-based modifications of gravity also typically require a new scale in addition of acceleration to pass Solar System constraints \cite{Desmond2024}.

For galaxy clusters, where the intra-cluster gas is the dominant baryonic component, the BTFR predicted by MOND can in principle be broadly translated into a baryonic mass ($M_b$)-temperature ($T$) relation, $M_b \propto T^2$. This slope of 2 is reasonably consistent with the data \cite{Sanders1994}, but the normalization predicted by MOND is completely off. In other words, MOND as presently formulated, fails in galaxy clusters, where there would remain a residual missing mass problem. This missing mass corresponds to a factor of a few globally, but it can rise to very high values in the central parts of clusters. This need for residual missing matter in MOND has long been known \citep[e.g.,][]{The,Gerbal,Sanders1999,Sanders2003,Pointecouteau,Natarajan,Angus2008,Ettori19,Kelleher}, and the gravitational lensing from the Bullet cluster \cite{clowe,angus} has additionally shown that this residual DM appears to behave in a collisionless way, namely, like usual DM. It would therefore be tempting to combine MOND with a form of (hot) non-baryonic dark matter, such as sterile neutrinos of a few eV, which would naturally be relevant only on galaxy cluster scales and above. Howewer, simulations of structure formation in this MOND-hot dark matter context have shown that it inevitably leads to a considerable overproduction of massive galaxy clusters \cite{angus2}, which severely reduces the appeal of the idea.  Another solution may come from the embedding of MOND in a covariant context. Indeed, the MOND behavior, {\it if} fundamentally true, should be the weak-field limit of a relativistic theory; such relativistic theories most often imply additional fields to the usual metric \citep[e.g.,][]{TeVeS,BSTV,BIMOND}, allowing to reproduce MOND in static configurations, whilst preserving the same relation between the lensing potential and gravitational potential as in GR. In recent versions of such relativistic theories, these additional fields can also play the role of DM on cosmic scales \cite{Skordis}: whether or not such fields could also help explain away the residual missing mass in galaxy clusters within such a theoretical framework is, however, an open question. Preliminary studies in this context have e.g. demonstrated a mass-dependent enhancement of the gravitational boost followed by an oscillation at large radii, quite different from the MOND non-relativistic limit for clusters \cite{Durakovic}. A related possibility would be that the $a_0$ scale is itself dependent on other quantities, such as the depth of the potential well which clearly separates galaxy clusters from galaxies \cite{emond}; however, this would not {\it a priori} explain why the residual missing mass would behave in a collisionless way in colliding clusters. Finally, the residual missing mass in the MOND context could simply indicate that MOND actually predicts additional baryonic mass (in a collisionless form) in the center of galaxy clusters, for instance cold gas clouds \cite{cbdm} in a multiphase intergalactic medium. In the latter case, one could expect that the distribution of the missing mass would more or less follow that of the observed hot gas in galaxy clusters.

Irrespective of the possible solution, or absence thereof,  we aim hereafter to check whether the residual missing mass profile of galaxy clusters in a MOND context has some universal behavior, and whether it correlates in some ways with the baryonic content of the clusters. Such a profile could then serve as a crucible for relativistic theories of MOND in galaxy clusters or any other tentative hypothesis regarding this residual missing mass in the MOND context. In Section~\ref{sec:theory}, we present the theoretical framework and the methods, before presenting the results based on the analysis of 16 clusters from the Cluster Lensing And Supernova survey with Hubble (CLASH) survey in Section~\ref{sec:results}, and concluding in Section~\ref{sec:concl}. In the Appendix, we analyze the residual mass in MOND using kinematics data and discuss why such data yield results which appear to be in tension with lensing.

\section{Theory and Methods} \label{sec:MOND}
\label{sec:theory}

\subsection{Gravitational lensing in MOND}

Our starting point is any covariant theory \citep[e.g.,][]{TeVeS,BSTV,BIMOND,Skordis} for which the gravitational lensing potential in the weak-field limit is $\Phi_{\rm{lens}} \equiv \Phi - \Psi = 2 \Phi$, where $\Phi$ is the gravitational potential governing dynamics, and $\Psi$ is the relativistic potential; in other words, theories for which $\Psi = -\Phi$ in the static weak-field limit. 

In the classical limit, for a large set of particles of mass $m_i$ moving in a gravitational field generated by the matter density distribution $\rho=\sum_i m_i \delta({\bf x} - {\bf x}_i)$, and described by the potential $\Phi$, we have the following action:
\begin{align}
& S = S_{\rm{matter}} + S_{\rm{gravity}}  \\ 
& S_{\rm{matter}} = \int d^{3}x dt \, \left( \frac{1}{2}\dot{{\bf x}}^2 - \Phi \right) \rho,
\end{align}
and in the Newtonian limit of GR, $S_{\rm gravity} = -(1/8 \pi G) \int d^3x \, dt (\nabla \Phi)^2$. There is a wide variety of ways to modify this classical gravitational action to get a MOND-like behaviour at low accelerations \cite{BM84,Milgrom2010}, sometimes only on certain scales depending on the number of other dimensioned constants than acceleration involved or the number of fields involved \cite{Milgrom2023a,Milgrom2023b}. It is common nowadays to rely on the so--called ``quasi-linear MOND'' formulation. This formulation makes the non-linear character of the theory transparent through the introduction of $\Phi_{\rm{N}}$ as an auxiliary potential, and is particularly convenient to implement in numerical simulations \citep[e.g.][]{Lughausen,Renaud,Nagesh}. It is important to note here that the conclusions of the present study, assuming spherical symmetry for galaxy clusters, will be largely independent of that particular choice of {\it theory}, apart from the fact that only the MOND acceleration constant should play a role (and should remain a constant). It might well not be the case in some recent relativistic formulations of MOND \citep[see, e.g.,][]{Durakovic}, but it is nevertheless the zeroth order expectation for any relativistic theory boiling down to MOND in the quasi-static weak field limit. The assumption of spherical symmetry, in and by itself, could slightly bias the results, since most known clusters are not strictly speaking spherically symmetric. In this sense, our analysis is not better (but also not worse) than those made within GR with such an assumption \citep[see, e.g., Ref.][for an assessment of the moderate effect of geometry on MOND predictions in the context of flat galaxy disks]{Famaey2012}. Let us also note that our main interest lies in the central parts of clusters where that approximation is reasonable at leading order, at least for clusters in equilibrium.

For the sake of clarity, we briefly remind the basics of the quasi-linear MOND formulation below. The total action is, as usual, a sum of two parts -- the matter and the gravity part, respectively -- but now we have:
\begin{align}
& S_{\rm{gravity}} = - \frac{1}{8 \pi G} \int d^{3}x dt \, \left( 2 \nabla \Phi \cdot \nabla \Phi_{\rm{N}} - a_{0}^2 Q(y^2) \right), 
\end{align}
where $y \equiv |\nabla \Phi_{\rm{N}}|/a_0$, and $Q(y^2)$ is in principle a non-linear function, which needs to satisfy certain requirements, as we will discuss below. Varying the matter action with respect to the matter density $\rho$, one recovers Newton's second law $\ddot{{\bf x}} = - \nabla \Phi$. Therefore, $\Phi$ plays the role of the total gravitational potential (to which matter couples), sourced by the total matter density $\rho$. Variation of the action with respect to the potential $\Phi$ yields the classical Poisson equation for baryonic matter as
\begin{align}
\nabla^2 \Phi_{\rm{N}} = 4 \pi G \rho. 
\label{eq:Poisson-Phi_N} 
\end{align}
The auxiliary gravitational potential $\Phi_{\rm{N}}$ is therefore just the usual Newtonian potential sourced by $\rho$, without any modification. The remaining step is the variation of the the gravity action with respect to $\Phi_{\rm{N}}$, which yields the modified Poisson equation for the total gravitational field to which matter actually couples,
\begin{equation}
\nabla^2 \Phi = \nabla \cdot \Big( Q'(y^2) \nabla \Phi_{\rm{N}}  \Big), 
\label{eq:Poisson-Phi} 
\end{equation}
with $Q'(y^2) \equiv \frac{d}{d y^2} Q(y^2)$. The non-linearity of the right-hand side of \eqref{eq:Poisson-Phi} is necessary to account for the ``dark-matter-like'' effects at galactic scales. The core idea of MOND is that 
\begin{equation*}
\nabla \Phi \rightarrow \nabla \Phi_N \; {\rm for} \; \nabla \Phi_N \gg a_0, \; {\rm and} 
\end{equation*}
\begin{equation}
\label{eq:milgrom}
\nabla \Phi \rightarrow \sqrt{a_0 \nabla \Phi_N} \; {\rm for} \; \nabla \Phi_N \ll a_0,
\end{equation}
therefore, the function $Q$ just needs to satisfy
\begin{equation}
Q(y^2) \rightarrow y^2 \; {\rm for} \; y^2 \gg 1 \; {\rm and} \; Q(y^2) \rightarrow \frac{4}{3}(y^2)^{3/4} \; {\rm for} \; y^2 \ll 1.
\end{equation}
In this paper, we will use the phenomenologically successful `simple' function for $Q'$ (see \citep{{Famaey2012}} and references therein), 
\begin{equation}
Q'(y^2) \!=\!\! \frac{1}{2} \left(1 + \left(1 + 4y^{-1}\right)^{\frac{1}{2}}\right).
\end{equation}

The general solution of Poisson equation is then equivalent to equation\eqref{eq:milgrom} in the asymptotic regimes, up to a curl field correction, and is precisely equal to it in a spherically symmetric system, which we will assume for galaxy clusters. 

In GR, we can relate the lensing potential and its source through 
\begin{equation}
\nabla^2 \Phi_{\rm{lens}} \equiv \nabla^2 (\Phi - \Psi) = 8 \pi G \rho_{\rm GR}, \label{eq:lensing-potential}
\end{equation}
where $\rho_{\rm GR}$ refers to any type of matter, baryonic or dark. Hence in spherical symmetry, we have for the enclosed mass
\begin{align}
M_{\rm{lensGR}}(r) \equiv \frac{r^2}{2G} \frac{\rm d}{{\rm d}r}\left( \Phi - \Psi \right) = \frac{r^2}{G} \frac{{\rm d} \Phi(r)}{{\rm d}r} \label{eq:lensing-mass}.
\end{align}
Lensing observations measure the combined potential $\Phi-\Psi$, which in turn can yield a reconstruction of $M_{\rm{lensGR}}$. To find the equivalent source mass for a spherically symmetric system in MOND, which we define as $M(r)$, one can use equation~(\ref{eq:Poisson-Phi}), which implies that $\nabla \Phi = Q'(y^2) \nabla \Phi_{N}$, and then simply solve for the enclosed mass $M(r)$ in 
\begin{equation}
M_{\rm{lensGR}}(r) = \frac{r^2}{G} Q'(y^2) \frac{{\rm d} \Phi_{\rm N}(r)}{{\rm d}r} \equiv  Q'(y^2) M(r).  
\label{main}
\end{equation}
Note that $M(r)< M_{\rm{lensGR}}(r)$, or in other words, the enclosed mass needed to produce a given lensing potential is smaller in MOND than in GR. Finally,
\begin{equation}
M(r) = 4 \pi \int dr \, r'^2 \rho(r'),
\label{eq:mass-definition}
\end{equation}
where $\rho$ is the matter density needed to reproduce the observations in MOND, part of which is certainly of baryonic nature, and the rest is the missing mass we are trying to reconstruct here. Note that the assumption of $\Psi = -\Phi$ for the (unspecified) relativistic MOND theory under consideration implies that observations of the galactic velocity field or the X-ray profile of the intracluster gas, providing measurements of the potential $\Phi$ as a function of distance from the cluster's center, will yield a dynamical mass equivalent to the lensing mass by construction, in MONDian gravity as well as in GR and Newtonian gravity, where $M_{\rm dynN} = M_{\rm{lensGR}}$. Importantly, since we assume spherical symmetry, the results of the present analysis are also valid for the aquadratic Lagrangian \cite{Bekenstein84} classical limit of modified gravity MOND, since the curl field correction to equation\eqref{eq:milgrom} then vanishes in that framework as well as in quasi-linear MOND. 

\subsection{Missing mass profiles in MOND}
\label{sec:missingmass}
In a first step, we will reconstruct the enclosed mass and density profiles, $M(r)$ and $\rho(r)$, needed to produce MONDian effects in accordance with lensing observations in galaxy clusters. In practice, assuming spherical symmetry, we rely in the following on the (parametric) estimate of $M_{\rm{lensGR}}(r)$ from lensing observations, and compute the enclosed mass profile needed to reproduce the same lensing signal in MOND, $M(r)$, which can be compared to the observed enclosed baryonic mass profile of the cluster. If MOND would not need any residual missing mass in clusters, the two quantities would be identical.

From the MONDian lensing mass $M(r)$, we then subtract the observed gas mass $M_\text{gas}(r)$, the stellar mass of the Brightest Cluster Galaxy (BCG) and companion galaxies, $M_{BCG}^*(r)$, which provide the dominant baryonic component at $r<0.05 \,\text{Mpc}$, and the mass of the other galaxies of the cluster $M_*(r)$. We do not consider the mass of the Intra-cluster Light (ICL), i.e. stars in the cluster region which are not associated to galaxies, since the contribution to the total stellar mass is small,
and is already partially accounted for, through the stellar mass of the BCG (see e.g. \cite{Contini2020,Furnell2021}). Then, at each radial bin $r$ we can define the missing mass up to radius $r$, and the associated density, as  
\begin{align}\label{eq:deltam}
& \delta M(r) = M(r)-M_\text{gas}(r)-M_*(r)-M_{BCG}^*(r)\,, \nonumber \\
& \rho_{\delta M}(r)=\frac{1}{4\pi r^2}\frac{\text{d} \left[ \delta M(r) \right]}{\text{d}r},
\end{align}
where $M(r)$ is the total mass predicted by MOND, $M_\text{gas}(r)$ is the contribution of the X-ray emitting hot gas of the ICM, $M_*(r)$ is the stellar mass in galaxies (except for the BCG), and $M_{BCG}^*(r)$ is the stellar mass of the BCG \footnote{The  profiles for the gas in our sample were provided by Vincenzo Salzano after request, and are computed following the prescription described in \cite{Laudato21}. They were double-checked in the Chandra Data Archive, and are provided in the file ClusterInfo.py at \url{https://zenodo.org/records/15299349}. We model the minor baryonic component $M_*(r)$ in our sample as $M_*(r)=M_\text{gas}(r)\times f_\text{gal}(r/r_{200})\,$, where $f_\text{gal}(x)$ is assumed to be the same for all clusters and normalized on our chosen reference cluster MACS~1206. We have further checked that ignoring the mass profiles of the galaxies beyond the BCG entirely does, in fact, not alter significantly any of the final outcomes of our analysis.}. 

It is important to notice here that the density profile above may become negative beyond a given radius, depending on the slope of the mass profile computed from equation~\eqref{main}. We will call that point the {\it turnaround radius} ($r_0$), and we will always truncate the profile beyond that point in order to preserve the physical character of the density if it is to be made of matter (see, however, \citep{Durakovic}). This feature is related, by construction, to the parametric form of the Newtonian/GR mass profile adopted to adjust the lensing observations, and we do not attempt here to determine whether it is an inescapable consequence of cluster lensing. Studying whether the lensing signal in the outskirts of clusters might be reconciled with the isolated MOND expectation, or whether the influence of the external gravitational field \cite{Kelleher} could influence those outskirts, or whether it would indicate a sharper drop as in \cite{Durakovic}, would need a fully non-parametric approach which should be the topic of further works. We can formally define the turnaround radius $r_0$ as the radius at which
\begin{equation}
\frac{\text{d} \left[ \delta M(r) \right]}{\text{d}r}\bigg|_{r=r_0} = 0.
\end{equation}

We are interested in understanding the scaling of the missing mass/density with radius, and the existence of possible universality features in it. In this regard, we will fit the missing mass with a generic density profile, namely, the ``$\alpha$-$\beta$-$\gamma$'' model (e.g., \citep{Zhao96}),
\begin{equation}
    \label{eq:abg}
    \rho_{\delta M}(r)=\frac{\rho_\text{s}}{\left(\frac{r}{r_\text{s}}\right)^\gamma \left[ 1 + \left(\frac{r}{r_\text{s}}\right)^\alpha\right]^{(\beta-\gamma)/\alpha}}\,,
\end{equation}
where $\rho_\text{s}$ and $r_\text{s}$ represent the characteristic density and scale radius of the profile, $\gamma$ represents the (logarithmic) inner slope, $\beta$ represents the outer slope, and $\alpha$ determines the sharpness of the transition between the inner and outer slopes, which we here fix to 1 and do not fit\footnote{This assumption is based on a preliminary analysis in which all parameters were treated as free variables. We observed no significant improvement in the fit when optimizing $\alpha$ alongside the other parameters, with very poor BIC. Furthermore, the best-fit values of $\alpha$ for our cluster sample are centered around $\langle \alpha \rangle = 1.02$. Given these preliminary results, we adopted the simpler $\beta-\gamma$ model with $\alpha$ fixed to 1.}. We will first fit all other parameters together for each cluster, and then fix the slopes in the search for possible global scaling relations between the characteristic density, scale radius and other observables. Finally, we will explore whether a ``dark mass-follows-gas'' profile -- which might be expected if the residual missing mass is made of missing baryons -- could equally well or even better represent the data.

\subsection{Lensing data}

Our analysis is based on a sample of 16 clusters from the CLASH survey, reconstructed through strong+weak lensing observations \cite{Umetsu2014, Umetsu2015}. For a review on weak-lensing of galaxy clusters we refer to \cite{Umetsu_review}. The CLASH sample is made of massive and (mostly) relaxed clusters, typically with an X-ray gas temperature larger than 5~keV and displaying a high degree of dynamical relaxation. No lensing data were used in the initial selection of these clusters to avoid any preferential alignment along the line of sight. A few clusters however do show some signs of departures from symmetric X-ray surface brightness distributions. In \cite{Umetsu16}, strong lensing, weak lensing shear, and magnification data for these clusters were combined by integrating data from Hubble observations and ground-based wide-field multicolor imaging, primarily using Suprime-Cam on the Subaru telescope. \cite{Umetsu16} fitted the lensing data of the CLASH sample by using a Navarro-Frenk-White profile \cite{navarro97} (NFW hereafter), providing the best fit parameters and the uncertainties, in which the relevant observational uncertainties were all taken care of. The error analysis accounts for cosmic noise (i.e cosmic shear acting as noise), the scatter in the lensing profile due to projection effects, as well as for observational uncertainties \cite{Umetsu_review}. It should be noticed that the latter modelling and fitting was performed in the context of General Relativity. Hereafter, the MCMC chains for the NFW fits are used together with the chains for the Burkert profile \cite{Burkert01} and the Hernquist profile \cite{Hernquist01} fits\footnote{The MCMC chains have been provided by Keiichi Umetsu as a private communication}.

\section{Results}
\label{sec:results}
\subsection{Residual missing mass profiles}\label{sec:resultlensing}

Given the lensing observations, we start from the mass profile, $M_{\rm lensGR}(r)$ for each cluster, and by virtue of equation~\eqref{main} we reconstruct the profile of the total ``baryonic'' mass needed in MOND, $M(r)$. To account for possible biases in the parametric modeling of the profiles, we consider three ansatz for the GR lensing mass profile, namely the NFW, the Burkert  and the Hernquist profiles, and we then average over them\footnote{We run also the lensing analysis assuming a generalized NFW (gNFW), characterized by an additional free slope parameter with respect to the NFW. However, the uncertainties in this case are extremely large, especially in the cluster core; moreover the additional parameter does not provide any significant improvement on the Posterior. Thus we discarded this model from the averaging procedure.}. We finally subtract the various observed baryonic contributions as explained in Section \ref{sec:missingmass}, see equation \eqref{eq:deltam}, to obtain the MOND residual missing mass profiles $\delta M(r)$. These are plotted for each individual cluster on Figure \ref{fig:missing_average}.

We then proceed considering the following cases for parametrizing the missing mass profiles:

\begin{figure*}
\centering
\includegraphics[width=0.8\textwidth]{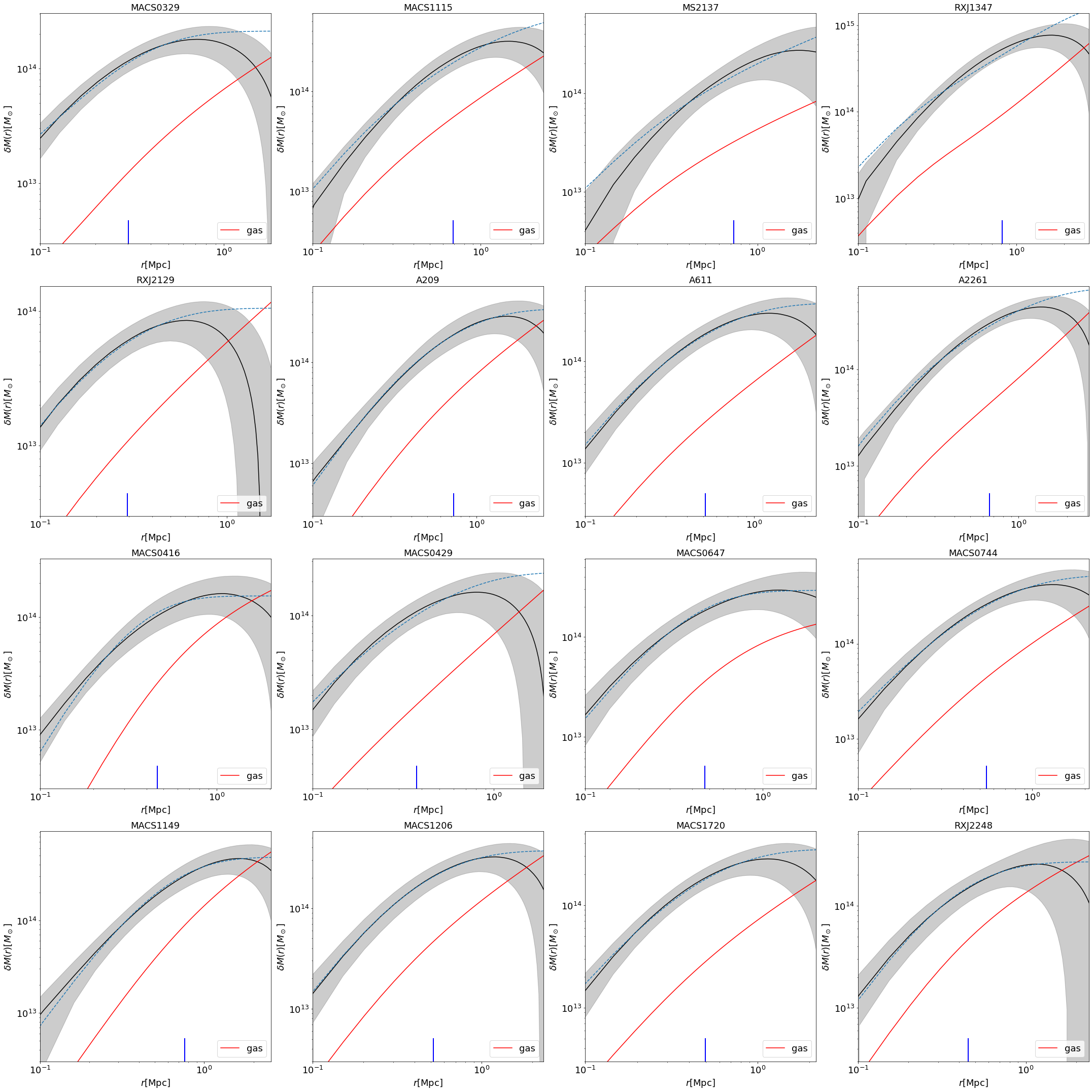}
\caption{MOND residual missing mass profiles $M(r)$ for all the galaxy clusters in the sample (black solid lines) together with their 68\% C.L. (gray-shaded areas).
The red solid lines indicate the gas mass profile following a beta-model (see the file ClusterInfo.py at \url{https://zenodo.org/records/15299349}). We overplot as blue-dashed curves the best fit from the `dark mass folows gas' case of equation \eqref{eq:propto}. The vertical blue ticks on the abscissa denote, on the other hand, the best fit scale radius $r_\text{s}$ with $\beta = 6$, from Sect. \ref{sec:resultlensing}, Case 2.}
\label{fig:missing_average}
\end{figure*}

\begin{itemize}
\item{\bf Case 1 ($\beta-\gamma$)}: We fit the missing mass $\delta M$ with the $\alpha$-$\beta$-$\gamma$ model of equation~\eqref{eq:abg} up to the turnaround radius of each cluster, fixing $\alpha = 1$ after our initial search. The main result of this fit is a very tight constraint on $\gamma$. Averaging over all clusters, we find an average value of 
\begin{equation}
\langle\gamma\rangle=0.015\,, \,\,
 \end{equation}
 and a weighted geometric mean $< 10^{-3}$, with very low scatter when accounting for the addition in quadrature of the scatter among the sample and the intrinsic uncertainties of the fit for each cluster. This means that the inner slope is very close to a core (which would correspond to $\gamma=0$). No relevant features have been found for $\gamma$ in relation to the gas mass or any other cluster property. This fitting procedure, however, does not yield a proper upper limit for the outer slope, for which the one sigma upper limit increases together with the size of the uniform prior. The lower limit however never goes below $\beta=3.7$, and the best-fit outer slope is typically extremely sharp ($\langle \beta \rangle \sim 6$), in line with the general trends previously reported by, e.g., \cite{Kelleher}. Concerning the two other parameters, the scale radius $r_\text{s}$ and characteristic density $\rho_\text{s}$, it is worth noticing that they exhibit a quite strong correlation, even though with large error bars due to the poor constraint on the upper limit of $\beta$. We will explore this correlation hereafter, once fixing all slopes. The main conclusion of this analysis is that the profiles are typically cored ($\gamma =0$), and have a sharp outer slope with $\beta > 3.7$.  

\begin{table*}
     \caption{Parameters of the MOND missing mass density profiles for the CLASH sample, adopting profiles with $\alpha=1$, $\gamma = 0$, and different values of $\beta$. The characteristic density $\rho_\text{s}$ is in units of ${\rm M}_\odot/{\rm Mpc}^3$ and the scale radius $r_\text{s}$ in Mpc. Column two, three: $\beta=4$. Column four, five: $\beta=5$. Column six, seven: $\beta=6$. Column eight, nine: $\beta=10$. Constraints are given within 1~$\sigma$}
     \small 
    \begin{tabular}{c|cc|cc|cc|cc}
        \hline
         &  \multicolumn{2}{c|}{$\beta = 4$} & \multicolumn{2}{c|}{ $\beta = 5$} & \multicolumn{2}{c|}{$\beta = 6$} & \multicolumn{2}{c}{$\beta = 10$} \\ 
        \cline{2-9}
        Cluster & Log $\rho_\text{s} $ & $r_\text{s}$ &  Log $\rho_\text{s} $ & $r_\text{s}$ & Log $\rho_\text{s} $ & $r_\text{s}$ & Log $\rho_\text{s} $ & $r_\text{s}$ \\ [0.2cm]
       \midrule	
        MACS 0329 & $ 16.48^{+  0.16}_{-  0.13}$ & $  0.14^{+  0.01}_{-  0.03}$ & $ 16.39\pm 0.14$ & $  0.22^{+  0.03}_{-  0.04}$ & $ 16.31^{+  0.13}_{-  0.13}$ & $  0.30^{+  0.03}_{-  0.05}$ & $ 16.24^{+  0.13}_{-  0.12}$ & $  0.61^{+  0.07}_{-  0.09}$  \\ [0.2cm]
        MACS 1115 & $ 15.60^{+  0.14}_{-  0.12}$ & $  0.35^{+  0.04}_{-  0.06}$ & $ 15.52^{+  0.13}_{-  0.12}$ & $  0.53^{+  0.06}_{-  0.08}$ & $ 15.49^{+  0.08}_{-  0.11}$ & $  0.69^{+  0.08}_{-  0.07}$ & $ 15.39^{+  0.08}_{-  0.14}$ & $  1.44^{+  0.19}_{-  0.15}$ \\ [0.2cm]
        MS 2137 & $ 15.41^{+  0.15}_{-  0.15}$ & $  0.38^{+  0.05}_{-  0.08}$ & $ 15.33^{+  0.15}_{-  0.13}$ & $  0.57^{+  0.07}_{-  0.11}$ & $ 15.32^{+  0.11}_{-  0.18}$ & $  0.73^{+  0.13}_{-  0.12}$ & $ 15.26^{+  0.07}_{-  0.16}$ & $  1.47^{+  0.24}_{-  0.15}$ \\ [0.2cm]
        RXJ1347 & $ 15.79^{+  0.12}_{-  0.10}$ & $  0.41^{+  0.04}_{-  0.06}$ & $ 15.71^{+  0.12}_{-  0.11}$ & $  0.61^{+  0.06}_{-  0.09}$ & $ 15.68^{+  0.09}_{-  0.13}$ & $  0.81^{+  0.10}_{-  0.09}$ & $ 15.63^{+  0.10}_{-  0.13}$ & $  1.60^{+  0.22}_{-  0.17}$\\ [0.2cm]
	RXJ2129 & $ 16.24 \pm 0.17 $ & $  0.14^{+  0.02}_{-  0.03}$ & $ 16.15\pm 0.15$ & $  0.21^{+  0.03}_{-  0.04}$ & $ 16.06^{+  0.18}_{-  0.15}$ & $  0.29^{+  0.04}_{-  0.06}$ & $ 16.05^{+  0.11}_{-  0.16}$ & $  0.55^{+  0.09}_{-  0.07}$\\    [0.2cm]    
	A209       & $ 15.50\pm 0.11$ & $  0.36^{+  0.04}_{-  0.05}$ & $15.45\pm 0.11$ &   $0.55^{+  0.06}_{-  0.07}$ & $ 15.38^{+  0.15}_{-  0.19}$ & $  0.72^{+  0.19}_{-  0.13}$ & $ 15.36^{+  0.08}_{-  0.13}$ & $  1.38^{+  0.15}_{-  0.15}$ \\[0.2cm]
	A611   & $ 15.98^{+  0.14}_{-  0.12}$ & $  0.25^{+  0.03}_{-  0.04}$ & $ 15.90^{+  0.15}_{-  0.11}$ &   $0.37^{+  0.03}_{-  0.06}$ & $ 15.82^{+  0.14}_{-  0.12}$ & $  0.52^{+  0.06}_{-  0.08}$ & $ 15.77^{+  0.10}_{-  0.12}$ & $  1.01^{+  0.12}_{-  0.10}$ \\ [0.2cm]
	A2261 & $ 15.86^{+  0.11}_{-  0.12}$ & $  0.32 \pm  0.04$ & $ 15.76^{+  0.10}_{-  0.10}$ &   $0.48^{+  0.05}_{-  0.06}$ & $ 15.70^{+  0.09}_{-  0.11}$ & $  0.66^{+  0.08}_{-  0.07}$ & $ 15.64^{+  0.08}_{-  0.10}$ & $  1.31^{+  0.13}_{-  0.12}$\\[0.2cm]
	MACS0416  & $ 15.82^{+  0.14}_{-  0.13}$ & $  0.23^{+  0.03}_{-  0.04}$ & $ 15.76^{+  0.12}_{-  0.13}$ &   $0.34^{+  0.04}_{-  0.05}$ & $ 15.70^{+  0.11}_{-  0.11}$ & $  0.46^{+  0.05}_{-  0.06}$ & $ 15.58^{+  0.07}_{-  0.12}$ & $  0.98^{+  0.13}_{-  0.08}$ \\ [0.2cm]
	MACS0429   & $ 16.11^{+  0.18}_{-  0.17}$ & $  0.19^{+  0.03}_{-  0.04}$ & $ 16.04^{+  0.16}_{-  0.15}$ & $  0.28^{+  0.04}_{-  0.05}$ & $ 16.00^{+  0.17}_{-  0.16}$ & $  0.38^{+  0.06}_{-  0.08}$ & $ 15.97^{+  0.13}_{-  0.16}$ & $  0.72^{+  0.11}_{-  0.11}$ \\ [0.2cm]
	MACS0647  & $ 16.12^{+  0.14}_{-  0.15}$ & $  0.21^{+  0.03}_{-  0.04}$ & $ 16.01^{+  0.15}_{-  0.15}$ & $  0.34^{+  0.04}_{-  0.06}$ & $ 15.93^{+  0.16}_{-  0.16}$ & $  0.47^{+  0.07}_{-  0.08}$ & $ 15.88^{+  0.14}_{-  0.12}$ & $  0.94^{+  0.10}_{-  0.13}$ \\ [0.2cm]
	MACS0744 & $ 16.01^{+  0.13}_{-  0.11}$ &   $0.27^{+  0.03}_{-  0.04}$ & $ 15.92\pm 0.12$ & $  0.41^{+  0.05}_{-  0.06}$ & $ 15.89^{+  0.11}_{-  0.11}$ & $  0.54^{+  0.07}_{-  0.07}$ & $ 15.78^{+  0.09}_{-  0.17}$ & $  1.14^{+  0.22}_{-  0.14}$ \\ [0.2cm]
	MACS1149  & $ 15.62\pm0.13$ & $  0.39^{+  0.05}_{-  0.06}$ & $ 15.53^{+  0.11}_{-  0.12}$ & $  0.59^{+  0.07}_{-  0.09}$  & $ 15.51^{+  0.10}_{-  0.10}$ & $  0.76^{+  0.08}_{-  0.08}$ & $ 15.51^{+  0.05}_{-  0.16}$ & $  1.43^{+  0.25}_{-  0.09}$ \\ [0.2cm]
	MACS1206 & $ 15.96^{+  0.17}_{-  0.14}$ & $  0.26^{+  0.03}_{-  0.05}$ & $ 15.87^{+  0.14}_{-  0.13}$ & $  0.40^{+  0.05}_{-  0.07}$ & $ 15.85^{+  0.13}_{-  0.15}$ & $  0.52^{+  0.08}_{-  0.08}$ & $ 15.77^{+  0.09}_{-  0.14}$ & $  1.07^{+  0.14}_{-  0.11}$ \\ [0.2cm]
	MACS1720   & $ 16.05^{+  0.13}_{-  0.14}$ &   $0.23^{+  0.03}_{-  0.04}$ & $ 15.94\pm 0.13$ & $  0.36^{+  0.04}_{-  0.06}$ & $ 15.86^{+  0.14}_{-  0.12}$ & $  0.50^{+  0.06}_{-  0.08}$ & $ 15.78^{+  0.08}_{-  0.15}$ & $  1.02^{+  0.16}_{-  0.09}$ \\ [0.2cm]
	RXJ2248  & $ 15.94^{+  0.13}_{-  0.17}$ &   $0.24^{+  0.04}_{-  0.04}$  & $ 15.88^{+  0.20}_{-  0.15}$ & $  0.36^{+  0.04}_{-  0.08}$ & $ 15.90^{+  0.16}_{-  0.13}$ & $  0.45^{+  0.05}_{-  0.08}$ & $ 15.81^{+  0.07}_{-  0.16}$ & $  0.93^{+  0.13}_{-  0.08}$ \\ [0.2cm]
        \hline
    \end{tabular}
    \label{table1}
\end{table*}

\item{\bf Case 2 (fixing all slopes)}: Given the results found above, we experiment with a general profile with a lower number of parameters for the fitting mass profile, keeping $\alpha =1$, and further fixing $\gamma = 0$ (cored inner profile), while considering separately the cases $\beta = 4, 5, 6, 10$ for the outer slope. In each case, only the characteristic (central) density and scale radius are fitted. This case gives a much more favourable Bayesian Information Criterion (BIC hereafter, see e.g. \cite{Mamon19}) than the previous one, due to the removal of one free parameter, and it is equivalent to the fits performed by \cite{Kelleher}, who found a very slight preference for $\beta=4$ based on hydrostatic equilibrium of hot gas. In our case, the resulting fits to the data also showed no strong statistical preference for any of the four values above. The BIC, however, gives a slight preference for $\beta=6$. 

In particular, we found an average $\Delta$BIC $= 0.42$ between $\beta = 4$ and $\beta = 6$, in favor of the latter, with a maximum difference of $0.85$, always below $1$. For the cases $\beta = 5$ and $\beta = 10$, we obtained an average $\Delta$BIC, of $0.08$ and $0.16$ with $\beta = 6$, still in favor of the latter, with a maximum difference of $0.23$ and $0.74$ respectively. The minimal difference with the results of \cite{Kelleher} should therefore not be over-interpreted as it is statistically irrelevant, and could also simply result from clusters spanning a slightly different range of parameters (total gas mass, temperature, etc.). On the other hand, when fixing $\beta=3$, the average $\Delta$BIC with $\beta=6$ rises to $1.7$, with a maximum value of $2.75$, and is therefore significant, in accordance with our findings from Case~1. The values of the fitted parameters for each individual cluster when fixing $\beta = 4, 5, 6, 10$ are all given in Table~\ref{table1}. We have also included, as a vertical tick on the abscissa, the fitted scale radius for the $\beta=6$ case in Figure~\ref{fig:missing_average}.

In Figure \ref{fig:scatterAB} we show, for all cases $\beta = 4, 5, 6$ and $10$, the central density versus the scale radius, and how the scale radius is correlated with the observed gas mass $M_\text{gas}$ (computed within the radius where the enclosed density in GR -- hence effective density in MOND -- is 200 times the critical density of the universe at the cluster redshift).
 The latter correlation is particularly significant  if we exclude the lowest gas-mass cluster from the sample (MS~2137). Concerning the correlation between the scale radius and the (log) characteristic density, already identified in Case~1, the slope of the relation steepens with increasing $\beta$. In particular, assuming a linear-log relation of the type $r_s = a {\rm log}(\rho_s) + b$,
$a= -0.11\pm 0.03,\, -0.38\pm 0.04,\, -0.51\pm 0.07, \, -1.07 \pm 0.14$ for $\beta = 4,5,6,10$, respectively. The physical interpretation of this anti-correlation is that the total amount of residual missing mass always remains, within an order of magnitude, within a few $\sim 10^{14}$ solar masses, but can be more or less concentrated, hence with a higher characteristic density when the scale radius is smaller. Moreover, the overall value of $r_\text{s}$ tends to grow with $\beta$: as $\beta$ increases, a larger $r_\text{s}$ is needed to compensate the faster decrease of the density profile. Conversely, the overall value of $\rho_\text{s}$ decreases with $\beta$.
 
 We also search the literature for signs of mild disequilibria, and find that half (8/16) of the sample displays such signs. Such disequilibria should not in principle affect the lensing mass: indeed, such clusters, displayed as black star symbols on Figure \ref{fig:scatterAB}, do not appear to follow different scaling relations than the other half of the sample.

As a final visual illustration of the parametrized fits performed hereinabove, we provide in Fig.~\ref{fig:A209}, as an example, a comparison of the best-fit profile for the missing mass in the cluster Abell~209, for the four values of $\beta$ considered here. This shows how close these different fits are to each other, and nicely illustrates how the scale radius increases with $\beta$.

\item{\bf Case 3 ($\beta$ as a function of $\gamma$)}: 
Finally, in order to check that the above results, especially on the central constant density core, are fully independent from the assumption of a fixed $\beta$, we now consider $\gamma$ as a free parameter with $\gamma \in [0, 0.6]$, and a linear dependence of $\beta$ on $\gamma$ according to $\beta = 4 +3\times\gamma$ or $\beta = 8 -5\times\gamma$. Fitting these two profiles to all clusters, we confirm that $\gamma \to 0$ is favored for all clusters, in both cases. These profiles actually give a BIC significantly worse than the models with $\alpha = 1, \gamma = 0$ and a fixed $\beta$ between 4 and 10 (due to the presence of an additional free parameter). The values of $\beta$ close to 4 and 8 found from those fits come from $\gamma$ being close to zero and from the fixed relation between $\beta$ and $\gamma$. As discussed earlier, for fixed values of the parameters, there is not statistically meaningful difference in the quality of the fits between $\beta=4$ and $\beta=5, 6$ or $10$. No relevant features have been found for $\gamma$ in relation to the gas mass.
\end{itemize}

\begin{figure*}
\centering

\includegraphics[width=0.6\textwidth]{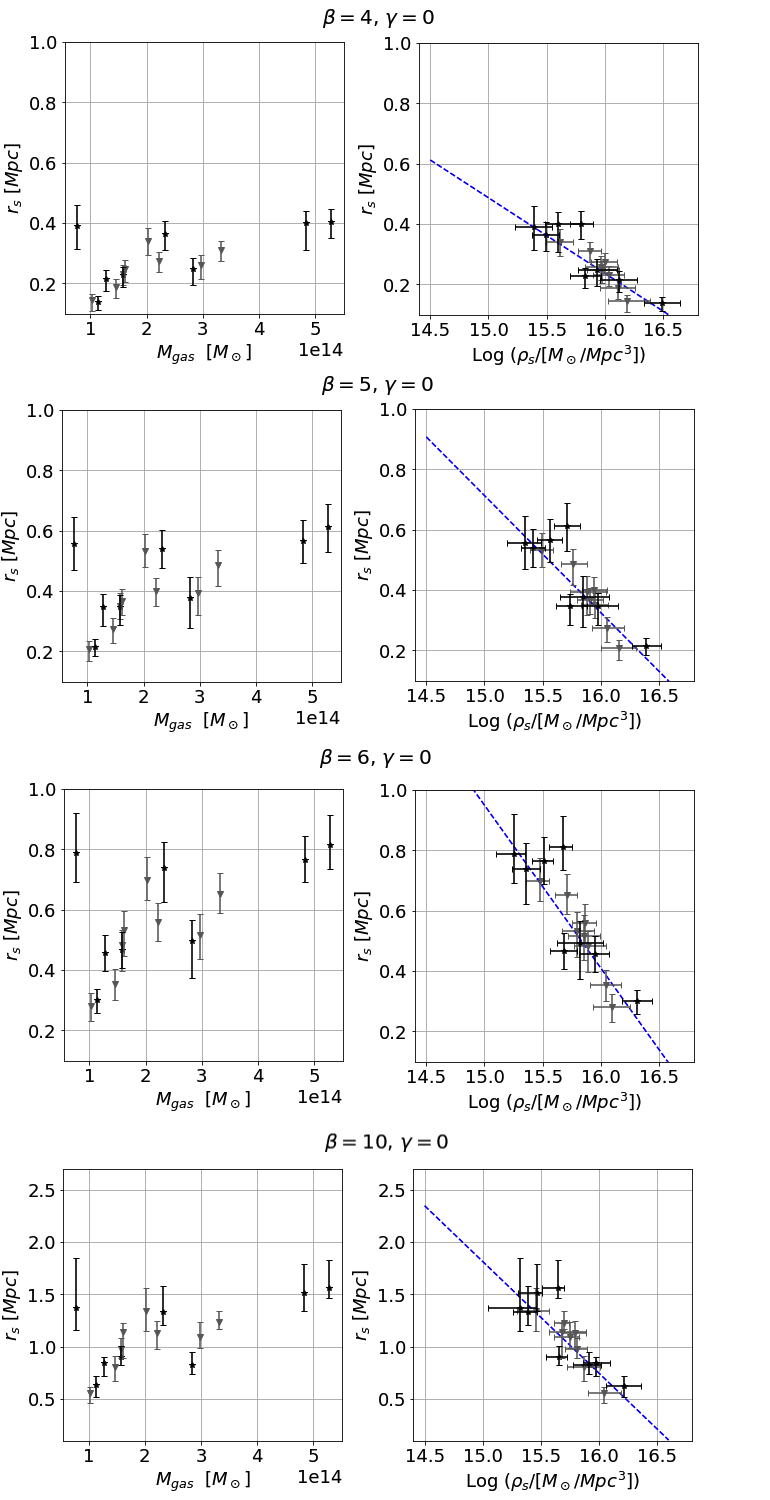}
\caption{Plots of the parameter $r_\text{s}$ vs the observed gas mass (left) and vs $\text{Log}\,\rho_\text{s}$; from top to bottom: for Case 2, with $\beta = 4,5,6,10$, as discussed in Section \ref{sec:resultlensing}. The error-bars represent the 68\% confidence intervals, while the blue dotted lines represent the best-fit linear regression of the $r_\text{s}$ vs $\text{Log}\,\rho_\text{s}$ relation. Black stars identify the clusters with signs of a disturbed state, while clusters at equilbrium are denoted by grey triangles. For more details on the analysis, see Section \ref{sec:results}.}
\label{fig:scatterAB}
\end{figure*}

\begin{figure}
\includegraphics[width=0.4\textwidth]{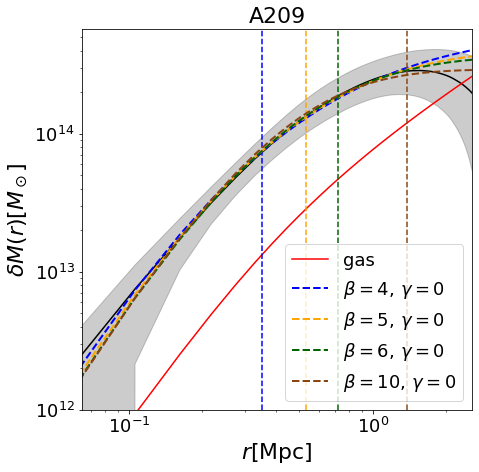}
\caption{MOND residual missing mass for Abell 209 (black solid line and gray-shaded area), together with the gas mass (red line) and the best fit profiles for Case 2, with $\beta = 4,5,6,10$, as discussed in Section \ref{sec:resultlensing}. The dashed vertical lines indicate the corresponding best fit scale radii $r_\text{s}$.}
\label{fig:A209}
\end{figure}

\subsection{Does the missing mass follow the gas density ?}

\begin{table*}
\centering
\caption{Parameters of the CLASH clusters: name (column 1), temperature (column 2, from \cite{Postman2012}), total gas mass (column 3), total MOND missing mass (column 4), 
missing-to-hot-gas density ratio (column 5), exponential cut-off radius (column 6), and flag for disequilibrium features (column 7, $*$ symbol if signs of disequilibria exist).}
\begin{tabular}{ccccccc}
Cluster & $kT$(keV) & $M_\text{gas}$($10^{14} {\rm M}_\odot$) & $\delta M(r_0)$($10^{14} {\rm M}_\odot$) & ${\rm log}(\eta)$ & $r_0/\lambda$(Mpc)& disequilibrium\\
\hline
\\
MACS 0329 & $8.0 \pm 0.5$  & $1.13$  & $1.79 \pm 0.47$ & $  1.27^{+  0.08}_{-  0.07}$ & $  0.24\pm 0.05$   & $*$ \cite{Girardi_2024} \\
MACS 1115 &  $8.0 \pm 0.4$ & $2.03$  & $3.17 \pm 1.09$  & $  0.66\pm0.06$ & $  1.40^{+  0.64}_{-  0.53}$   & \\
MS 2137 &  $5.9 \pm 0.3$  & $0.77$  & $2.73 \pm 1.60$ & $  0.70^{+  0.03}_{-  0.04}$ & $  4.14^{+  3.00}_{-  0.76}$ & $*$ \cite{Chiu_2012} \\
RXJ1347 & $15.5 \pm 0.6$  & $5.28$ & $7.71 \pm 2.37$ & $  0.81 \pm  0.05$ & $  1.61^{+  0.57}_{-  0.48}$   &   $*$ \cite{Johnson2012} \\
RXJ2129 &  $5.8 \pm 0.4$ & $1.03$  & $0.85 \pm 0.27$ & $  1.06^{+  0.11}_{-  0.10}$ & $  0.23\pm  0.07$   \\
A209 & $7.3 \pm 0.5$ & $2.33$ & $2.88 \pm 1.07$ & $0.91^{+  0.08}_{-  0.07}$ & $ 0.60^{+  0.14}_{-  0.13}$  & * \cite{Feng_2024} \\
A611 & $7.90 \pm 0.35$ & $1.62$ & $2.99 \pm 1.02$ & $  1.12 \pm  0.07$ & $  0.48^{+  0.12}_{-  0.11}$ \\
A2261 & $7.6 \pm 0.3$ & $3.32$ &  $4.46 \pm 1.10$ & $  1.05^{+  0.06}_{-  0.05}$ & $  0.65 \pm 0.14$   &  \\
MACS 0416 & $ 7.5 \pm 0.8$ & $1.58$ & $1.61 \pm 0.61$ & $  1.28^{+  0.12}_{-  0.11}$ & $  0.19\pm  0.04$  & * \cite{Raney21} \\
MACS 0429 & $6.00 \pm 0.44$ & $1.46$ & $1.61 \pm 0.62$ & $  1.03\pm  0.08$ & $  0.38^{+  0.12}_{-  0.10}$  &  \\
MACS 0647 & $13.3 \pm 1.8$ & $1.28$ & $2.95 \pm 1.24$ & $  1.07^{+  0.11}_{-  0.09}$ & $  0.35 \pm  0.09$  & * \cite{Giovannini21} \\
MACS 0744 & $8.9 \pm 0.8$ & $2.22$ & $4.18 \pm 1.45$ & $  1.01^{+  0.06}_{-  0.07}$ & $  0.55^{+  0.15}_{-  0.11}$  &  \\
MACS 1149 & $8.7 \pm 0.9$ & $4.84$
& $4.65 \pm 1.65$ & $ 1.07^{+  0.09}_{-  0.10}$ & $  0.38^{+  0.09}_{-  0.06}$ &  * \cite{Giovannini21}\\
MACS 1206 & $10.8 \pm 0.6$  & $2.97$ & $3.24 \pm 1.03$ & $  0.98^{+  0.07}_{-  0.08}$ & $  0.40^{+  0.10}_{-  0.07}$  &  \\
MACS 1720 &  $6.6 \pm 0.4$ &  $1.58$ & $2.83 \pm 0.96$ & $  1.08\pm 0.06$ & $  0.46 \pm 0.09$ &  \\
RXJ2248 & $12.4 \pm 0.6$ & $2.83$ & $2.55 \pm 1.20$ & $  0.81^{+  0.12}_{-  0.10}$ & $  0.39^{+  0.13}_{-  0.11}$  & * \cite{Mercurio2021} \\
\end{tabular}
\label{table2}
\end{table*}

So far, we have re-constructed the missing mass $\delta M$ in a systematic way, without caring as to what its underlying nature could be. We found, in accordance with the literature, that the generic missing mass profile is close to following a cored density profile in the central parts followed by a sharp decline with a power-law slope sharper than 3.5 in the outskirts ($r \gtrsim 0.5$~Mpc). If this missing mass would be baryonic, for instance in the form of cold gas clouds, then one might perhaps expect a close correlation with the distribution of the hot gas in the central parts of the cluster. Visually, it indeed appears in Figure \ref{fig:missing_average} that $\delta M$ seems to roughly scale with that of the enclosed gas mass in the central parts.

To check for this possible behaviour in a more quantitative manner, we first consider a case similar to those of the previous sections, but where the $\gamma$ parameter, instead of being fixed to zero, is fixed to have the same central slope as the gas profile. The other parameters are fixed to be $\alpha=1$ (as always) and $\beta=6$. We find that the average $\Delta$BIC$=0.5$ compared to the $\beta=6$, $\gamma=0$ case remains mostly reasonable, with only one case reaching a $\Delta$BIC of 2. To push our investigations further, we then write down an effective model for the missing mass taking into account the hot gas mass profile of the cluster under consideration,

\begin{equation}
\label{eq:propto}
\rho_{\delta M}(r) = \eta \cdot \rho_\text{gas}(r) e^{-\lambda r/r_0},
\end{equation}
where $\eta$ is a constant indicating the ratio of residual missing mass density over hot gas density. Obviously, the gas mass density does not fall down as sharply as the missing mass density in the outskirts, which can be modelled thanks to the exponential cut-off factor depending on the free parameter $\lambda$ which ensures a smooth and sharp decrease of $\rho_{\delta M}(r)$. The relevant physical parameter of the profile fitted here is actually the cut-off radius $r_0/\lambda$. For each individual cluster, the best-fit profile is shown as a blue dashed curve in Figure \ref{fig:missing_average}. Interestingly, this model actually provides an overall good BIC compared to the other cases presented above. The only two exceptions are the clusters MS~2137 and RXJ~1347, for which $\Delta$BIC$ \simeq 7$ in favour of the case with fixed $\gamma=0$, $\beta=6$. The best-fit parameters for all clusters of the sample with this profile are listed in Table~\ref{table2}, together with a flag for clusters displaying signs of disequilibria.

\begin{figure*}
\centering
\includegraphics[width=0.99\textwidth]{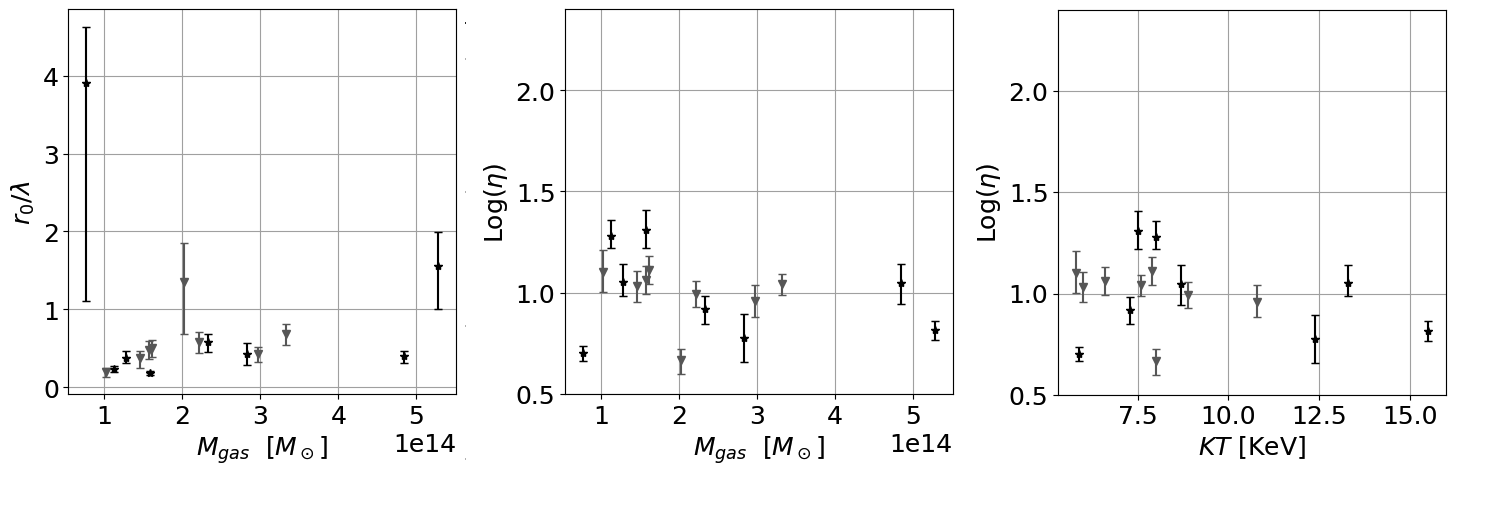}
\caption{Left and central plots: fitted parameters in the effective missing density model of equation \eqref{eq:propto} as a function of the gas mass of the cluster sample. Right: $\eta$ (in log) as a function of the cluster temperatures as found in \cite{Postman2012}. For a detailed discussion of these results, see Section \ref{sec:results}. Note the remarkable constancy of both $\eta$ and $r_0/\lambda$ with both gas mass and temperature. As above, black stars identify the clusters with signs of lack of equilibrium.
}
\label{fig:gasMas_ModelAdvanced}
\end{figure*}

Figure \ref{fig:gasMas_ModelAdvanced} illustrates the remarkable uniformity for the missing-to-hot-gas density ratio and cut-off radius, which are almost constant, with weighted mean $ \langle {\rm log} \, \eta  \rangle= 0.93$ and geometric weighted mean $\langle r_0/\lambda \rangle= 0.43$~Mpc. The latter parameter is a bit more scattered than the missing-to-hot-gas density ratio, but the largest values are also associated with large error bars, compatible at 1$\sigma$ with values closer to the average. On the other hand, one can note a slight tendency for the missing-to-hot-gas density ratio to increase with decreasing gas mass, although this tendency should be confirmed with a larger sample. As in the previous cases, clusters displaying signs of disequilibria do not display a different behaviour than the other ones, indicating that the lensing results are not sensitive to them.

\section{Discussion and conclusions}
\label{sec:concl}
We have used parametric mass profiles derived from strong and weak-lensing shear, as well as magnification data, for a sample of 16 clusters from the CLASH survey, in order to compute the equivalent mass profile needed to produce the same gravitational field in MOND. This has allowed us to study possible scalings of the MOND residual missing mass in the inner parts with the observed baryonic distribution in galaxy clusters.

Fitting a very general profile to the residual missing mass profiles, we found that its inner slope is very close to a core whilst the outer slope is sharp with power-law slopes larger than 3.5, in accordance with the results of \cite{Kelleher} from the hydrostatic equilibrium of hot gas in the X-COP sample. We also note a strong anti-correlation between the characteristic density of the residual missing mass and its scale radius, whose normalization and slope depends on the outer slope, whilst the scale radius itself tends to correlate with the total hot gas mass of  clusters. 

Pushing further the study of a possible connection between the distribution of baryons and that of the missing mass, we also consider the possibility of a ``dark mass-follows-gas'' profile in the central parts, followed by an exponential cut-off. We show that such a model provides a particularly good description of the missing profiles, with a remarkable uniformity of the missing-to-hot-gas density ratio and cut-off radius. We, however, caution that this uniformity is obtained from lensing alone, but that cross-checks from hydrostatic equilibrium of X-ray emitting gas, and from the kinematics of galaxies, should also be performed to reach a firm conclusion on the typical values. In the Appendix, we perform such an analysis based on the kinematics data of the cluster MACS~1206, which confirms that the scaling of the observed gas mass profile and residual missing mass profile works excellently, but we caution that it yields significantly smaller values of the the missing-to-hot-gas density ratio in the central parts ($\text{log}\,\eta = 0.33$) and a larger cut-off radius $r_0/\lambda = 2.12$~Mpc). Clearly, the lensing and kinematics results are in tension. This inconsistency between kinematics and lensing data is inherent to the data, and not to the gravity theory, and most plausibly due to systematic uncertainties in the interpretation of kinematics data (similar to the hydrostatic bias), despite the fact we used the most accurate available dataset on the market. The kinematics results, therefore, appear to be biased with respect to lensing. We also notice that, before identifying the cause of the discrepancy accurately, including more kinematics  datasets in our analysis would then not improve the picture, but act as to strengthen the bias.

Since our study relies on a parametric form of the lensing potential, we do not attempt here to study whether the lensing signal is in tension with MOND in the outskirts of the clusters, which would require a fully non-parametric approach that should be the topic of further works. We emphasize though that the lensing reconstruction of the total mass profile of the cluster has been performed for different mass profile parametrizations. It is worth noting that \cite{Kelleher} have shown that the external field effect expected in MOND (and mostly irrelevant for our present study of the mass profile in the inner parts) could have a significant effect, whilst \cite{Durakovic} have shown that relativistic versions of MOND might also induce a sharp drop of the gravitational field in the outskirts.
Our results in the inner parts of clusters can serve as a crucible for such relativistic theories of MOND, or for any other tentative hypothesis regarding the nature of the clusters residual missing mass in the MOND context. For instance, the enhancement of the gravitational boost followed by an oscillation at large radii found by \cite{Durakovic} could provide a satisfactory explanation only if the boost is confined to the central parts of the cluster and following the hot gas mass profile, which is not in qualitative agreement with the theoretical profiles presented in  \cite{Durakovic}. It is worth noting that the same CLASH clusters have previously been studied in the context of the Radial Acceleration Relation by \cite{Tian}, who concluded that a relation similar to that found in galaxies could be deduced from lensing data, but with a characteristic acceleration an order of magnitude larger than expected in MOND. This would appear to be in line with a variation of the $a_0$ scale as a function of other quantities such as the depth of the potential well \cite{emond}. This also might appear consistent at zeroth order with our results. However this is not precisely equivalent to our present suggestion. Indeed, an $a_0$ rescaling would imply that the residual missing mass would be entirely correlated (in the usual MONDian way) to the total enclosed baryonic mass in the central parts, including the BCG mass. Our results, on the contrary, exhibit an excellent correlation with the distribution of the gas mass alone. 
In the X-COP clusters of \cite{Kelleher}, it also appears clearly that the acceleration profile generated by observable baryons does not follow the shape of the acceleration generated by the residual missing mass. It would, however, be most worthwhile to check whether our findings could be corroborated in these X-COP clusters, namely if removing the influence of the BCG would result in a scaling similar to that found in our CLASH sample. This appears, by eye, to be the case but with typically smaller values of the missing-to-hot-gas density ratio (see also the Appendix hereafter), which might be related to the hydrostatic bias.

In summary, our study uncovers for the first time a close connection between the distribution of X-ray emitting hot gas and that of the residual missing mass of MOND in galaxy clusters, which can serve as a guide for testing and excluding tentative solutions to the problem of MOND in galaxy clusters. We nevertheless caution that, while the suggested connection between the hot gas mass profile and the missing mass profile appears to hold also from a kinematics analysis (see Appendix), the exact values of the missing-to-hot-gas density ratio can be smaller than reported in our lensing analysis. 

\vspace{5 mm}

\acknowledgements
The authors wish to warmly thank the anonymous referee as well as Tobias Mistele for constructive feedback and extremely helpful checks. BF acknowledges funding from the European Research Council (ERC) under the European Union's Horizon 2020 research and innovation program (grant agreement No. 834148) and from the Agence Nationale de la Recherche (ANR projects ANR-18-CE31-0006 and ANR-19-CE31-0017). IDS acknowledges funding by the Czech Grant Agency (GA\^CR) under the grant number 21-16583M. LP acknowledges funding from the Czech Academy of Sciences under the grant number LQ100102101. The statistical analysis was designed and led by BF and IDS, the Python code was developed by LP under the supervision of the former two, the lensing data used in the analysis were provided by K. Umetsu, and the gas-profile data were provided by V. Salzano respectively. A reproduction package for the statistical analysis of this work in Python is available at: \url{https://zenodo.org/records/15299349}

\bibliography{sample,moreReferences}

\vspace{5 mm}

\appendix \section{Kinematics of the cluster MACS 1206}
\label{sec:kinematics}
 As a complementary investigation to the above lensing analysis, we have compared the results with those obtained from kinematics data in the cluster MACS 1206. As we also commented briefly in the main text, these results appear to be in tension with those derived from lensing. This is independent of the gravity theory, and likely due to inherent systematics despite the good accuracy of the MACS 1206 kinematics. Therefore, caution is needed in their interpretation and comparison with the lensing analysis. The MACS 1206 cluster has been extensively studied in the spectroscopic follow-up campaign of CLASH with the VIMOS spectrograph at Very Large Telescope (CLASH-VLT \cite{Rosati2014}), complemented with observation from the MUSE integrated field instrument for the cluster core \cite{Caminha2017}. Data about the stellar velocity dispersion of the BCG itself have also been collected (see \cite{Biviano23c} and references therein\footnote{The public redshift catalog, references and strong lensing models for this clusters are available at \href{https://sites.google.com/site/vltclashpublic/data-release?authuser=0}{the CLASH-VLT website}}). The galaxy masses mass are taken from \cite{Annunziatella14}. The reconstruction of the total dynamical mass follows the method presented in \cite{Biviano23b}, where an upgraded version of the \textsc{MG-MAMPOSSt} code \cite{Pizzuti2021,Pizzuti2023a} has been applied to fit the member galaxies velocity field, along with the velocity dispersion of the BCG. Note that in kinematics' mass determination of clusters, one has to account for an additional unknown, namely the velocity anisotropy profile $\beta = 1 - (\sigma^2_\theta +\sigma^2_\phi)/(2\sigma^2_r)$, 
where $\sigma^2_{r,\theta,\phi}$ are the velocity dispersion components in the radial, tangential and azimuthal direction, respectively. The velocity anisotropy profile $\beta(r)$ is modeled using a ``generalized Tiret'' \citep[gT,][]{Tiret} and a ``generalized Osipkov-Merritt'' (gOM) models \cite{Mamon19}, which account for a broad range of orbits phenomenology in clusters. These are applied to four different mass profile ansatz, the generalized NFW (gNFW) model, the Einasto profile \cite{Einasto65} with the characteristic exponent $m$ fixed to $m=3$, as well as the Burkert and Hernquist models already mentioned in Section~\ref{sec:resultlensing}. As done for the lensing analyses, we then average over all the mass and anisotropy profiles in the kinematics analysis, compute the equivalent MOND mass profile $M(r)$, and subtract all the baryonic components (i.e. the BCG, gas and galaxy profiles) to obtain $\delta M(r)$, according to equation \eqref{eq:deltam}. We find a total residual missing mass of $(3.10 \pm 0.63) \times 10^{14} {\rm M}_\odot$, only slightly lower than for the lensing analysis, but significantly lower in the most central regions. As already pointed out, this inconsistency is likely due to systematic uncertainties inherent to kinematics data, leading to biased results compared to lensing. Another reason for the discrepancy might however be linked to the fact that the kinematics analysis excludes from the phase-space a background group of galaxies located in the projection very near the cluster core, which is perhaps affecting in the lensing mass reconstruction. Importantly, the model of equation~\eqref{eq:propto} is still a very good match to the data, but the main difference with the lensing analysis is that  $\text{log}\,\eta = 0.33^{+0.04}_{-0.05}$, much lower than in the lensing analysis. To compensate for a similar total amount of residual missing mass, the value of $r_0/\lambda = 2.12 \pm 0.45$~Mpc is on the other hand much larger than what was found by the lensing analysis.

\end{document}